# Crystallization of metallic glass as a grain-boundary nucleated process: experimental and theoretical evidence for the grain structure of metallic glasses

Nikolay V. Alekseechkin

Akhiezer Institute for Theoretical Physics, National Science Center "Kharkiv Institute of Physics and Technology", 1, Akademichna Str., Kharkiv 61108, Ukraine
Email: n.alex@kipt.kharkov.ua

Experimental DSC and Avrami curves for the crystallization of metallic glasses demonstrate nucleation at grain boundaries and thus indicate their grain structure, which refutes the generally accepted idea of glass as a homogeneous "frozen" liquid obtained as a result of avoiding crystallization. Under certain conditions, this nucleation mechanism results in the appearance of two-peak DSC curves and three-step Avrami plots which are observed experimentally. To clarify these conditions, isothermal and non-isothermal surface-nucleated crystallization of a spherical particle is considered within the framework of a simple analytical model of nucleation and growth.

# 1. Introduction

Thermal analysis (DSC[1] and DTA measurements) of phase transformations is a powerful experimental tool for understanding these processes; in combination with appropriate theoretical studies, it can provide insight into the mechanism of phase transformation and make it possible to evaluate some important characteristics of the process. The heat flow $w$ accompanying the phase transformation and measured in the experiment (the DSC curve) is proportional to the rate of change of the volume fraction (VF) $X(t)$ of the transformed material: $w \sim dX(t)/dt = \beta dX/dT$, where $T$ is temperature and $\beta = dT/dt$ is the heating rate in non-isothermal experiments. The VF $X(t)$ itself is then obtained as the relative area under the DSC curve. The aim of the theory is to obtain dependences $X(t)$ for various mechanisms of the nucleation and growth of a new phase.

The classical Kolmogorov-Johnson-Mehl-Avrami (KJMA) [1-4] theory was the first such theory; it is widely used to interpret experimental results on the crystallization of metallic glasses. The KJMA theory assumes homogeneous nucleation (equal probability of the appearance of a nucleus at any point in the system) and a linear law of growth of the appearing nuclei (the case of interface-controlled growth). Although the case of a parabolic growth law (diffusion-controlled growth) is beyond the scope of Kolmogorov's model [1], it has been shown [5] that it is also described with good accuracy by the KJMA equation.

However, experimental results in many cases are not consistent with the KJMA theory. The first evidence of a non-KJMA type of transformation is the DSC curves having the appearance of double peaks (which the authors usually call "overlapping peaks"), when an additional peak or "shoulder" appears on the ascending branch of the main peak [6-15], Fig. 1a; these peaks can also be well separated from each other. This shape of the DSC curves apparently indicates *nucleation at grain boundaries*; previously, such a shape was observed in the process of spontaneous amorphization of a metastable crystalline phase [16] and was reproduced in Ref. [17] within the framework of a theoretical model, where an equation for the non-isothermal DSC curve was derived. In this process, the amorphous phase nucleates at the grain boundaries of the crystalline phase. Nuclei grow into the grain body and simultaneously cover the parent boundary. At some time, the boundaries turn out to be completely transformed; Cahn [18] called this effect "boundary surface saturation" (BSS). After this moment, nucleation no longer occurs and one-dimensional growth of the new phase occurs. The first peak or shoulder on the DSC curve corresponds to the stage of transformation of the boundary surface, while the main peak describes the mentioned one-dimensional growth.

---

[1] **Abbreviations**: VF volume fraction; GBN grain-boundary nucleated; BSS boundary surface saturation; DSC differential scanning calorimetry; DTA differential thermal analysis; KJMA Kolmogorov-Johnson-Mehl-Avrami.

Thus, although the transformation proceeds continuously, nevertheless, the grain-boundary nucleated (GBN) process appears to be a two-stage process if BSS occurs at an early stage; the two peaks reflect this fact. The shoulder mentioned above is obviously an "incomplete" peak. To obtain a peak instead of a shoulder, a model of anisotropic growth with tangential (along the boundary), $u_\tau$, and normal, $u_n$, growth rates of the nucleus was used in Ref. [17]; $u_\tau > u_n$ due to the fact that the mobility of atoms on the grain boundary is higher than in the grain body and therefore the BSS effect is more pronounced in this case.

The second experimental evidence of the non-KJMA type of crystallization of metallic glasses is the Avrami plots [$\ln(-\ln(1-X))$ vs. $\ln t$] consisting of three sections [19], Fig. 1b. This form of the Avrami plot is obtained within the framework of a recently proposed new model of GBN transformation [20, 21] which uses closed domain geometry – spherical [20] and flat [21] geometries - for grain boundaries instead of the Cahn model of random planes [18]. It was shown in Ref. [20] that the first section of this three-step curve corresponds to the KJMA transformation mode at an early stage of the process. The most important and informative is the final section of the curve – a steep bend; it corresponds to the one-dimensional *radial* growth of the crystalline phase after BSS. In Ref. [20], it was shown that the Avrami exponent (which is the slope of the Avrami plot) increases sharply over time for this growth regime. Thus, it is this section of the Avrami plot that allows us to recognize the GBN type of transformation; in Cahn's model it is absent. The second section of the Avrami curve corresponds to the transition between the two indicated regimes. Sometimes the first (KJMA) section can be skipped during measurements, e. g., if the KJMA mode occurs at the VFs $X < \bar{X}_1$, where $\bar{X}_1$ is the lower experimental limit for VF measurements ($\bar{X}_1 = 0.05$ is usually reported in the literature); in this case, the Avrami plot consists of two parts (curve 3 in Fig. 1b).

Thus, the experimental results considered indicate the grain structure of metallic glasses, in contrast to the generally accepted concept of glass as a homogeneous "frozen" liquid obtained as a result of avoiding crystallization. If metallic glass were homogeneous, the crystalline nucleus would appear with equal probability at any point in it, and the Avrami plot for crystallization would always be a straight KJMA line.

Ubbelohde [22] viewed the process of glass formation as the result of competition between crystalline and non-crystalline clusters in a supercooled liquid. As a result of the predominance of non-crystalline clusters, the liquid transforms into a polycluster amorphous body [23]; the kinetic model of this process is considered in Ref. [24].

In experiments, a domain structure of some glasses [25] and the polycluster structure of a Zr-based bulk metallic glass [26] were observed. It was found that the structure of the clusters is also inhomogeneous: they include compact nanometer atomic complexes – subclusters measuring 1.5-3 nm with a high atomic packing density [26]; the previously observed regions of medium-range ordering in metallic glasses [27] have the same sizes. The structure of small atomic complexes and local order in glasses are the subject of intensive research [28-35]. The more components in the alloy, the more types of local order, which facilitates cluster growth. It is well known that multicomponent alloys with atoms of different sizes are good glass formers.

In the mentioned Zr-based metallic glass, internal interfaces (intercluster boundaries) with a density of $10^6$ cm$^{-1}$ were observed [26]. Atoms in the intercluster boundary layer have a binding energy 0.13-0.43 eV lower than inside the cluster [26]. Thus, intercluster boundaries are preferred nucleation sites, so that the crystallization of metallic glasses proceeds as a GBN transformation. In the experiment, the surface of intercluster boundaries could be revealed by contrasting structures in the following way. Under certain conditions indicated below (e. g., high nucleation rates), BSS occurs at an early stage. If at this point the crystallization is interrupted (the sample is quenched), then we will obtain an amorphous sample with a network of crystalline film inside which represents the surface of intercluster boundaries.

The new model of GBN transformations [20] is based on the surface-nucleated transformation of a spherical particle. The latter demonstrates the above-described main features of GBN transformations, so a simple spherical particle model is used here as an illustrative model.

## 2. Analytical model for nucleation and growth

For demonstrative calculations, some analytical model of nucleation and growth is needed; VF equations include temperature-dependent nucleation and growth rates. We assume that these temperature dependences are qualitatively the same in one-component and multicomponent systems and therefore the standard one-component model will give qualitatively correct results. Indeed, both nucleation and growth are thermally activated processes, so the corresponding rates depend on temperature in both mentioned systems mainly according to an exponential law. For the parameters included in the thermodynamic and kinetic equations, some characteristic values are adopted to obtain the required crystallization temperature range and the typical average crystal grain size $\overline{R}_{gr}$ after the completion of crystallization.

For the difference in chemical potentials between the glassy and crystalline phases, a quadratic expansion is used near the temperature $T_*$,

$$\Delta\mu(T) = \Delta\mu_0 - \Delta s_0(T - T_*) - \frac{\Delta c_p^0}{2T_*}(T - T_*)^2 \qquad (1)$$

where $\Delta\mu_0 = 0.8kT_*$, $\Delta s_0 = 0.4k$, and $\Delta c_p^0 = 0.7k$ are the differences in chemical potentials, entropies, and heat capacities at the temperature $T_* = 700\,°K$, $k$ is the Boltzmann constant. In a multicomponent alloy, this is the chemical potential $\overline{\mu}$ averaged over the components which can be defined using the Gibbs free energy as follows:

$$G = \sum_i N_i \mu_i = N_{tot} \sum_i x_i \mu_i = N_{tot} \overline{\mu}, \quad \overline{\mu} = \sum_i x_i \mu_i, \quad x_i = N_i / N_{tot} \qquad (2)$$

The Gibbs formula for the nucleation work, $W = (4\pi/3)\sigma R_c^2$, after substituting $R_c = 2\sigma a^3/\Delta\mu$ [36, 37] for the critical radius, gives

$$W(T) = \frac{16\pi}{3} \frac{\sigma^3 a^6}{[\Delta\mu(T)]^2} \qquad (3)$$

where $a^3$ is the mean atomic volume and $a = 3 \times 10^{-8}$ cm is the mean interatomic distance. The surface tension $\sigma = 120$ dyne/cm is assumed to be constant here for simplicity, although it is also depends on temperature. In GBN transformations, heterogeneous nucleation mechanism occurs, so that Eq. (3) must be multiplied by the function of contact angle [38]. However, in our model this function can be taken into account by appropriately changing the above parameters.

The kinetics of phase transformation is largely determined by the frequency of atomic jumps

$$K_D(T) = \nu e^{-\frac{g}{kT}} \qquad (4)$$

where $\nu = 10^{13}\,s^{-1}$ is the characteristic frequency of atomic vibrations and $g = 2.7 \times 10^{-12}$ erg (1.7 ev) is the activation energy of self-diffusion.

The nucleation rate equation [39] can be reduced to the following form:

$$I(T) = 2N\nu \sqrt{\frac{\sigma a^2}{kT}} e^{-\frac{g + W(T)}{kT}} \qquad (5)$$

where $N = a^{-3}$ is the number of atoms per unit volume.

The interface-controlled growth rate is given by the equation [38]

$$u(T) = aK_D(T)\left[1 - e^{-\frac{\Delta\mu(T)}{kT}}\right] \quad (6)$$

while the diffusion-controlled growth rate is

$$\frac{dR}{dt} = \frac{c_d D(T)}{2R}, \quad D(T) = a^2 K_D(T) \quad (7a)$$

from where

$$R(t',t) = \sqrt{c_d}\left(\int_{t'}^{t} D(\tau)d\tau\right)^{1/2} \quad (7b)$$

where $D(T)$ is the diffusion coefficient and $c_d$ is a constant taken equal to one in calculations; $R(t',t)$ is the radius at time $t$ of the nucleus that appeared at time $t'$. Since diffusion growth is slower than linear growth, a smaller value of $g = 1.7 \times 10^{-12}$ erg together with $\sigma = 128$ dyne/cm was taken for numerical calculations. In a multicomponent alloy, kinetic processes are determined by the slowest component [37], therefore $g$ is the activation energy for this component.

The parameters used give $\bar{R}_{gr} \approx 1.3 \mu m$ for the mean grain "radius" in the isothermal KJMA crystallization. The non-isothermal Kolmogorov VF equation has the form [1]

$$X(t) = 1 - \exp\left[-\int_0^t I(t')V(t',t)dt'\right], \quad V(t',t) = \frac{4\pi}{3}R^3(t',t) \quad (8)$$

When heating at a constant rate $\beta$ from the initial temperature $T_0 = 500\ °K$,

$$T(t) = T_0 + \beta t, \quad t = \frac{T - T_0}{\beta} \quad (9)$$

so that $R(t',t) = \beta^{-\xi} r(T',T)$ and

$$r(T',T) = \begin{cases} \displaystyle\int_{T'}^{T} u(T'')dT'' & \text{for linear growth} \\ \sqrt{c_d}\left(\displaystyle\int_{T'}^{T} D(T'')dT''\right)^{1/2} & \text{for diffusion growth} \end{cases} \quad (10)$$

when we go from $t$ to $T$ in non-isothermal VF equations; $\xi = 1$ and $1/2$ for linear and diffusion growth, respectively. Although the linear growth law is realized only in the isothermal case, for brevity we use the term "linear" for interface-controlled growth in the non-isothermal case as well.

Fig. 2 shows the non-isothermal KJMA kinetic curves $X(T)$, Eq. (8), together with the corresponding rates of VF change $dX/dT$.

## 2. Isothermal surface-nucleated crystallization of a spherical particle

Isothermal equation for the VF $Q(t) = 1 - X(t)$ in the process of the surface-nucleated transformation of a spherical particle of radius $R_0$ has the following form for linear growth [20]:

$$Q(\tau, \alpha_s) = \begin{cases} (1-\tau)^3 + 3\int_{1-\tau}^{1} e^{-\alpha_s \phi_1(x,\tau)} x^2 dx, & \tau < 1 \\ 3\left\{ \int_0^{\tau-1} e^{-\alpha_s \phi_2(x,\tau)} x^2 dx + \int_{\tau-1}^{1} e^{-\alpha_s \phi_1(x,\tau)} x^2 dx \right\}, & 1 \leq \tau \leq 2 \\ 3\int_0^1 e^{-\alpha_s \phi_2(x,\tau)} x^2 dx, & \tau > 2 \end{cases} \quad (11)$$

where $\tau = ut/R_0$ is the dimensionless time;

$$\phi_1(x,\tau) = \frac{2(1-x)^3 - 3\tau(1-x)^2 + \tau^3}{x}, \quad \phi_2(x,\tau) = 12(\tau - 1 - \frac{1}{3}x^2),$$

and

$$\alpha_s(T) = \frac{\pi}{3} \frac{I_s(T)}{u(T)} R_0^3 \quad (12)$$

is a key parameter that determines the nature of the transformation. Here, $I_s(T) = I(T)/\omega$ is the surface nucleation rate related to the bulk nucleation rate $I(T)$ through the density of grain boundaries $\omega$ (the area of grain boundaries per unit volume); $\omega = 3/R_0$ for a spherical particle [20]. Defining by $t^* = R_0/u$ the time corresponding to $\tau = 1$, we have $\alpha_s = I_s S_0 t^*/12$, $S_0 = 4\pi R_0^2$. Thus, this parameter is proportional to the mean number of nuclei formed on the boundary surface during time $t^*$ (including fictitious nuclei [2]).

The boundary of the spherical particle is transformed according to the KJMA 2D law [20] at $\tau < 2$:

$$X_s(\tau) = 1 - e^{\alpha_s \tau^3} \quad (13)$$

For diffusion growth, a similar VF equation holds [20], but with replacement $\tau \to \sqrt{\tau}$, $\tau = c_d Dt/R_0^2$, other functions $\phi_i(x,\tau)$ and $\alpha_s^{(d)}(T) = \pi(I_s(T)/c_d D(T))R_0^4$.

Fig. 3 shows that the parameter $\alpha_s$ depends significantly on temperature. At first, it increases due to an increase in the nucleation rate with temperature, then it decreases due to a decrease in the nucleation rate at higher temperatures (the function $I(T)$ has a maximum) and an increase in the growth rate (the functions $u(T)$ and $D(T)$ increase monotonically). Sufficiently large values of $R_0$ ($5 \times 10^{-3}$ cm and $2 \times 10^{-3}$ cm for linear and diffusion growth, respectively) were taken to get $\alpha_s = 10^4 \div 10^5$ in the working range.

Fig. 4a shows the Avrami plots for Eq. (11) for different values of the parameter $\alpha_s$. As can be seen, the shape of these curves differs significantly for small and large values of $\alpha_s$. For large values of $\alpha_s$, of the order of and greater than $10^2$, a three-step curve is formed, which is observed in the experiment. The expansion of $Q(\tau, \alpha_s)$ at $\tau \to 0$ shows [20] that the first section of this curve corresponds to the KJMA crystallization regime at the beginning of the process. The last section of this curve (the steep bend) corresponds to the one-dimensional radial growth of the crystalline phase after BSS, Fig. 4b, which is described by the power term in Eq. (11):

$$X_{1D}(\tau(t)) = 1 - (1-\tau)^3 = 1 - \left(1 - \frac{ut}{R_0}\right)^3 \tag{14a}$$

$$\frac{dX_{1D}(t)}{dt} = 3\left(1 - \frac{ut}{R_0}\right)^2 \frac{u}{R_0} \tag{14b}$$

The Avrami exponent for the function $(1-\tau)^3$ increases infinitely [20] at $\tau \to 1$, which corresponds to the completion of crystallization at large values of $\alpha_s$, Fig. 4b.

For small values of $\alpha_s$, of the order of and less than 1, the slope of the Avrami plot decreases to unity after the initial KJMA regime. However, this behavior is not associated with the growth of the crystalline phase; it simply reflects the property of the particle's finiteness: $Q(\tau(t), \alpha_s) \sim \exp(-12\alpha_s \tau) = \exp(-S_0 I_s t)$ at long times [20]. In other words, for the crystallization of a particle at long times, the appearance of at least one nucleus on its boundary is necessary; the probability that this event will not occur is determined by the mentioned exponent. For large values of $\alpha_s$, this effect does not appear, since in this case crystallization is completed at $\tau \sim 1$.

Considering the grain structure consisting of identical spherical particles, Fig. 5a, the new model of GBN transformations allows for the possibility of growing nuclei crossing grain boundaries. The VF equation for such a structure has the form of Eq. (11), but with two functions $[\phi_i(x,\tau) + c\phi_i^{(b)}(x,\tau)]$, $c = \omega R_0$, where the functions $\phi_i^{(b)}(x,\tau)$ take into account the contribution to the VF from the boundaries external to the given particle [20]. In other words, nuclei appearing at the outer boundaries can invade and transform a given particle, along with nuclei appearing at its own boundary, Fig. 5a. As a result, the Avrami plots for this structure are the same as for an isolated particle at large values of $\alpha_s$ and are qualitatively different at small values of $\alpha_s$, Fig. 5b.

For small values of $\alpha_s$, the Avrami plots are straight KJMA lines. Small values of $\alpha_s$ correspond to low $I_s$/small $R_0$/high $u$, Eq. (12). As the cell size $R_0$ in this network of boundaries decreases, the space for nucleation becomes more homogeneous, i.e. the KJMA model is approached. A low nucleation rate $I_s$ (and also $I$) means rare nucleation events randomly scattered over space, as in the KJMA model, while a high growth rate $u$ prevents the BSS effect. This explains the closeness of GBN crystallization kinetics to KJMA kinetics at small values of $\alpha_s$. Accordingly, the kinetic curves $X(\tau)$ and the DSC curves have a "normal" shape similar to the KJMA one, Fig. 5c. This is a confusing case for analyzing experimental data when the GBN process is perceived as a KJMA process.

At high values of $\alpha_s$, BSS occurs at an early stage of transformation; the boundary of a given particle is covered with a crystalline film (Fig. 4b) which prevents the penetration of nuclei from the outside. Thus, in this case, the influence of external boundaries is negligible and crystallization occurs as for an isolated particle (cf. Figs. 4a and 5b for $\alpha_s > 10^2$). The kinetic and DSC curves in Fig. 5c change their shape, approaching the corresponding asymptotic lines; the most part of the $X(\tau)$ curve is a power law (in contrast to the sigmoid shape for small values of $\alpha_s$), while the $dX/d\tau$ curve has a sharp maximum and a parabolic descending branch. In a real grain structure there is a distribution of grain sizes. This distribution "stretches" the Avrami plot in time and makes its third section shorter [20], as on the experimental curves (cf. Fig. 1b). The parameter $\alpha_s$ is given by the same Eq. (12), but with the mean grain size $\overline{R}_0$.

Thus, in the case of large values of $\alpha_s$ (below, only this case is considered), the spherical particle model is sufficient for understanding the main qualitative features of GBN crystallization; moreover, only the first branch of Eq. (11) (for $\tau < 1$) is sufficient. It is this case that allows us to recognize the GBN transformation in the experiment.

The Avrami exponent $n(t) = d\ln(-\ln Q(t))/d\ln t$ for the first branch of Eq. (11) is shown in Fig. 6 for different temperatures. It decreases with time from the initial KJMA value and increases sharply at the end of the transformation, as noted above; this sharp increase corresponds to the steep bend in the Avrami curve. When plotted as functions of the VF $X$, $n(X) = n(t(X))$, these curves become close to each other and partially coincide. Thus, the behavior of the function $n(X)$ exhibits a certain universality: the value of the Avrami exponent is determined mainly by the stage of the process, i.e. the value of $X$. A similar dependence $n(X)$ for the crystallization of metallic glass was obtained experimentally [40].

Fig. 7 shows isothermal DSC curves $dX/dt$ for the same conditions as in Fig. 6. Eq. (13) allows us to estimate the time $t_{BSS}$ of BSS from the equation $X_s(t_{BSS}) = 0.999$; these moments are shown by vertical lines. As can be seen, after $t_{BSS}$ the curve gradually passes to its asymptotic form given by Eq. (14b). Thus, an isothermal DSC curve has only one maximum occurring shortly before BSS, but its descending branch has a characteristic parabolic shape for linear growth, in contrast to the KJMA exponential shape, Fig. 2.

## 3. Non-isothermal surface-nucleated crystallization of a spherical particle

Moving on to non-isothermal crystallization of a spherical particle, we can use only the first branch of the non-isothermal VF equation [20], as shown above:

$$Q(t) = \left(1 - \frac{R_m(t)}{R_0}\right)^3 + \frac{1}{V_0} \int_{R_0 - R_m(t)}^{R_0} \exp\left[-\int_0^{t_m(\rho,t)} I_s(t')S(\rho;t',t)dt'\right](4\pi\rho^2)d\rho \tag{15a}$$

where $V_0 = (4\pi/3)R_0^3$, $R_m(t) = R(0,t)$, $S(\rho;t',t) = \pi R_0[R^2(t',t) - (R_0 - \rho)^2]/\rho$, and $t_m(\rho,t)$ is found from the equation $R(t_m,t) = R_0 - \rho$.

Moving from time to temperature, we get

$$Q(T,\beta) = \left(1 - \frac{\beta^{-\xi}r(T_0,T)}{R_0}\right)^3 + \frac{1}{V_0} \int_{R_0 - \beta^{-\xi}r(T_0,T)}^{R_0} \exp\left[-\beta^{-1}\int_{T_0}^{T_m(\beta,\rho,T)} I_s(T')S(\beta,\rho;T',T)dT'\right](4\pi\rho^2)d\rho$$

(15b)

where $S(\beta,\rho;T',T) = \pi R_0[\beta^{-2\xi}r^2(T',T) - (R_0 - \rho)^2]$ and $T_m(\beta,\rho,T)$ is found from the equation $\beta^{-\xi}r(T_m,T) = R_0 - \rho$. As in Eq. (10), the parameter $\xi$ here covers the cases of linear and diffusion growth; the corresponding expression from Eq. (10) should be taken for $r(T',T)$.

The first term describes non-isothermal one-dimensional radial growth:

$$X_{1D}(T,\beta) = 1 - \left(1 - \frac{\beta^{-\xi}r(T_0,T)}{R_0}\right)^3 \tag{16}$$

Its derivative proportional to the DSC curve at the corresponding stage is equal to

$$\frac{dX_{1D}(T,\beta)}{dT} = \frac{3\beta^{-1}}{R_0}\left[1 - \frac{\beta^{-1}r(T_0,T)}{R_0}\right]^2 u(T) \tag{17}$$

for linear growth and

$$\frac{dX_{1D}(T,\beta)}{dT} = \frac{3a}{2R_0}\sqrt{\frac{c_d}{\beta}}\left[1 - \frac{\beta^{-1/2}r(T_0,T)}{R_0}\right]^2 \left(\int_{T_0}^T K_D(T')dT'\right)^{-1/2} K_D(T) \tag{18}$$

for diffusion growth.

Finally, the non-isothermal counterpart of Eq. (13) has the form

$$X_s(T,\beta) = 1 - \exp\left[-\pi\beta^{-(2\xi+1)}\int_{T_0}^T I_s(T')r^2(T',T)dT'\right] \tag{19}$$

Fig. 8 shows the VF $X(T,\beta) = 1 - Q(T,\beta)$ and its derivative for different heating rates $\beta$. As can be seen, in the case of linear growth, a shoulder appears on the ascending branch of the DSC curve, as on the experimental curves (cf. Fig. 1a). In the case of diffusion growth, a shoulder appears on the descending branch at small values of $\beta$ which at higher values of $\beta$ turns into a second peak.

A more detailed analysis of these curves is presented in Fig. 9, where the auxiliary curves given by Eqs. (16)-(19) are added to explain this shape of the DSC curves. As can be seen, the maximum on the curve $dX/dT$ in Fig. 9a is the maximum of the function $dX_{1D}(T,\beta)/dT$, Eq. (17), which is the product of increasing ($u(T)$) and decreasing (brackets) functions. As before, we determine the temperature $T_{BSS}$ at which BSS occurs by the equation $X_s(T_{BSS}) = 0.999$ using Eq. (19). The shoulder is formed shortly before BSS, at $X_s \approx 0.9$, as are the first peaks in Fig. 9b. From this, we can conclude that the second (integral) term in Eq. (15b) is responsible for the appearance of the shoulder, as well as the peaks mentioned in Fig. 9b. This term contains the surface area $S$, hence the appearance of the shoulder and the corresponding peaks is a consequence of the BSS effect. However, the second important conclusion is that non-isothermality also plays a key role here: the shoulder or the corresponding peak can only appear due to the fact that the DSC curve after it continues to rise due to the increasing growth rate $u(T)$. This fundamentally distinguishes the non-isothermal case from the isothermal one, where BSS results in a single peak on the DSC curve, Fig. 7. Previously, the role of non-isothermality in the formation of two-peak DSC curves was noted in Ref. [17].

The initial (KJMA) portion of the $dX/dT$ curve is also shown in Fig. 9a. It should be emphasized that the higher the $\alpha_s$ value, the shorter the KJMA stage; e.g., for $\alpha_s \sim 10^5$ this is the case for $X < 0.01$.

As noted above, an additional peak associated with the BSS effect appears on the DSC curve in cases of special growth laws, when the boundary surface is transformed as if "separately" from the grain body, so that the process becomes quasi-two-stage. The model of Ref. [17] uses disk-shaped nuclei that grow faster along the boundary than along the normal to it. Fig. 9b shows that the diffusion growth law satisfies this criterion and a peak associated with the BSS effect appears on the DSC curve. As can be seen, BSS for the second curve (with two peaks) occurs at an early stage, $X \approx 0.2$. Since nucleation occurs only at the boundary surface, the latter is transformed predominantly by small nuclei having a high growth rate according to Eq. (7a). At the same time, the growth rate of the nucleus slows down as its size increases. Thus, a quasi-two-stage process is naturally realized here. The second peak is broad due to the slow diffusion growth of large nuclei; as in the previous case, it is formed due to the competition between the increasing and decreasing functions in Eq. (18).

## 4. Conclusions

1. The nature of the GBN crystallization process is determined by the parameter $\alpha_s$. At small values of $\alpha_s$ (of the order of and less than 1), GBN crystallization looks like KJMA one and therefore cannot be identified experimentally.

2. Only in the case of large values of $\alpha_s$ (of the order of $10^2$ and greater), GBN crystallization manifests itself in the experiment.

3. The first such manifestation is the three-step Avrami curve ending in a sharp bend. This bend corresponds to the one-dimensional radial growth of the crystalline phase after BSS. The dependence of the Avrami exponent on the VF, $n(X)$, has a characteristic "U-shaped" form.

4. The second manifestation is a shoulder or an additional peak on the *non-isothermal* DSC curve; both precede the mentioned one-dimensional radial growth. They represent a transformation of the boundary surface itself which is accompanied by the exhaustion of places for nucleation and the loss of two degrees of freedom for growth. The shoulder and the additional peak are a consequence of both the BSS effect and non-isothermality.

5. In such growth modes, when the boundary surface is transformed as if "separately" from the grains, a peak appears instead of a shoulder, i.e. the quasi-two-stage nature of the crystallization process becomes visible.

6. It should be emphasized that the Avrami plot is defined for isothermal processes, so the two mentioned manifestations – the three-step Avrami plot and the two-peak DSC curve – refer to different experimental conditions – isothermal and non-isothermal, respectively. The analogue of the Avrami plot for non-isothermal processes is the Ozawa plot [41, 20] – [$\ln(-\ln(1-X))$ vs. $\ln \beta^{-1}$] at a fixed $T$.

7. If GBN crystallization at large values of $\alpha_s$ is interrupted at the moment of BSS (the sample is quenched), then we obtain an amorphous sample with a network of crystalline film inside which represents the boundary surface.

## References

1. English translation: A. N. Kolmogorov, On the statistical theory of metal crystallization. Izv. Akad. Nauk SSSR Ser. Mat. **3,** 355-360 (1937), in: A. N. Shiryayev (Ed.), Selected Works of A. N. Kolmogorov. Mathematics and Its Applications (Soviet Series) (Springer, Dordrecht, 1992), **26**, pp. 188-192.

2. W. A. Johnson, R. F. Mehl, Reaction kinetics in processes of nucleation and growth. Trans. AIME **135,** 416-458 (1939).

3. K. Barmak, A commentary on: "Reaction kinetics in processes of nucleation and growth". Metallurgical and Materials Transactions A, **41A,** 2711-2775 (2010).

4. M. Avrami, Kinetics of phase change. II Transformation-time relations for random distribution of nuclei. J. Chem. Phys. **8,** 212-224 (1940).

5. N. V. Alekseechkin, Extension of the Kolmogorov-Johnson-Mehl-Avrami theory to growth laws of diffusion type. J. Non-Cryst. Solids **357**, 3159-3167 (2011).

6. H. E. Khalifa, K. S. Vecchio, Thermal stability and crystallization phenomena of low cost Ti-based bulk metallic glass. J. Non-Cryst. Solids **357,** 3393-3398 (2011).

7. K. M. Cole, D. W. Kirk, C. V. Singh, S. J. Thorpe, Role of niobium and oxygen concentration on glass forming ability and crystallization behavior of Zr-Ni-Al-Cu-Nb bulk metallic glasses with low copper concentration. J. Non-Cryst. Solids **445-446,** 88-94 (2016).

8. B. S. Murty, D. H. Ping, K. Hono, A. Inoue, Influence of oxygen on the crystallization behavior of $Zr_{65}Cu_{27.5}Al_{7.5}$ and $Zr_{66.7}Cu_{33.3}$ metallic glasses. Acta Mater. **48,** 3985-3996 (2000).

9. A. A. Tsarkov, E. N. Zanaeva, A. Yu. Churyumov, S. V. Ketov, D. V. Louzguine-Luzgin, Crystallization kinetics of Mg-Cu-Yb-Ca-Ag metallic glasses. Mater. Charact. **111,** 75-80 (2016).

10. J. Zhang, B. Shen, Z. Zhang, Crystallization behaviors of FeSiBPMo bulk metallic glasses, J. Non-Cryst. Solids **360,** 31-35 (2013).

11. Z. Zheng, J. Wang, Q. Xing, Z. Sun, Y. Wang, Effects of La addition on glass formation, crystallization and corrosion behavior of Gd-Al-based alloys. J. Non-Cryst. Solids **379,** 54-59 (2013).

12. B. Schwarz, U. Vainio, N. Mattern, S. W. Sohn, S. Oswald, D. H. Kim, J. Eckert, Combined in-situ SAXS/WAXS and HRTEM study on crystallization of $(Cu_{60}Co_{40})_{1-x}Zr_x$ metallic glasses. J. Non-Cryst. Solids **357,** 1538-1546 (2011).

13. X. Gu, L. Q. Xing, T. C. Hufnagel, Glass-forming ability and crystallization of bulk metallic glass $(Hf_xZr_{1-x})_{52.5}Cu_{17.9}Ni_{14.6}Al_{10}Ti_5$. J. Non-Cryst. Solids **311,** 77-82 (2002).

14. R. Wang, L. Shi, Y. Wu, J. Jia, Y. Shao, K. Yao, Effect of Mo on the glass forming ability and properties of Fe-B-C-P-Si-Mo bulk metallic glasses. J. Non-Cryst. Solids **629,** 122868 (2024).
15. X. Yang, T. Wang, Q. Li, R. Wu, B. Gao, Q. Yang, Effect of Cu content on crystallization behavior, mechanical and soft magnetic properties of $Fe_{80-x}Cu_xP_{13}C_7$ bulk metallic glasses. J. Non-Cryst. Solids **546,** 120274 (2020).
16. E. G. Ponyatovsky, O. I. Barkalov, Pressure-induced amorphous phases. Mater. Sci. Rep. **8,** 147-191 (1992).
17. N. V. Alekseechkin, A.S. Bakai, C. Abromeit, On the kinetics of spontaneous amorphization of a metastable crystalline phase. J. Phys.: Condens. Matter **13,** 7223-7236 (2001).
18. J. W. Cahn, The kinetics of grain boundary nucleated reactions. Acta Metall. **4,** 449-459 (1956).
19. L. Q. Xing, J. Eckert, W. Löser, L. Schultz, D. M. Herlach, Crystallization behavior and nanocrystalline microstructure evolution of a $Zr_{57}Cu_{20}Al_{10}Ni_8Ti_5$ bulk amorphous alloy. Phil. Mag. A **79,** 1095-1108 (1999).
20. N. V. Alekseechkin, Kinetics of the surface-nucleated transformation of spherical particles and new model for grain-boundary nucleated transformations. Acta Mater. **201,** 114-130 (2020).
21. N. V. Alekseechkin, Kinetics of grain-boundary nucleated transformations in rectangular geometries and one paradox relating to Cahn's model. Acta Mater. **221,** 117350 (2021).
22. A. R. Ubbelohde, Melting and Crystal Structure (Clarendon Press: Oxford, 1965).
23. A. S. Bakai, The polycluster concept of amorphous solids, in: H.-J. Güntherodt, H. Beck (Eds.), Glassy Metals III (Springer: Berlin, 1994) pp. 209-255.
24. N. V. Alekseechkin, A. S. Bakai, N. P. Lazarev, Kinetics of hardening of a supercooled liquid with concurrent formation of several solid phases. Fiz. Nizk. Temp. **21,** 565-575 (1995). English translation: Low Temp. Phys. **21,** 440-448 (1995).
25. J. C. Phillips, The physics of glass. Phys. Today **35**(2), 27 (1982).
26. A. S. Bakai, I. M. Mikhailovskij, T. I. Mazilova, N. Wanderka, Field emission microscopy of the cluster and subcluster structure of a Zr-Ti-Cu-Ni-Be bulk metallic glass. Low Temp. Phys. **28,** 279 (2002).
27. Y. Hirotsu, M. Uehara, M. Ueno, Microcrystalline domains in amorphous Pd77.5Cu6Si16.5 alloys studied by high resolution electron microscopy. J. Appl. Phys. **59,** 3081-3086 (1986).
28. R. Wang, Short-range structure for amorphous intertransition metal alloys. Nature **278,** 700-704 (1979).
29. A. Hirata, P. Guan, T. Fujita, Y. Hirotsu, A. Inoue, A. R. Yavari, T. Sakurai, M. Chen, Direct

observation of local atomic order in a metallic glass. Nature Mater. **10,** 28-33 (2011).

30. D. B. Miracle, A structural model for metallic glasses. Nature Mater. **3,** 697-702 (2004).
31. D. B. Miracle, Efficient local packing in metallic glasses. J. Non-Cryst. Solids **342,** 89-96 (2004).
32. O. N. Senkov, D. B. Miracle, Effect of the atomic size distribution on glass forming ability of amorphous metallic alloys. Mater. Res. Bulletin **36,** 2183-2198 (2001).
33. D. B. Miracle, The efficient cluster packing model – An atomic structural model for metallic glasses. Acta Mater. **54,** 4317-4336 (2006).
34. Y. Q. Cheng, E. Ma, H. W. Sheng, Atomic level structure in multicomponent bulk metallic glass. Phys. Rev. Lett. **102,** 245501 (2009).
35. K. J. Laws, D. B. Miracle, M. Ferry, A predictive structural model for bulk metallic glasses. Nature Communications **6,** 8123 (2015).
36. T. Philippe, D. Blavette, P. W. Voorhees, Critical nucleus composition in a multicomponent system. J. Chem. Phys. **141,** 124306 (2014).
37. N. V. Alekseechkin, Thermodynamics and kinetics of nucleation in binary solutions. Chem. Phys. **517,** 138-154 (2019).
38. J. W. Christian, The Theory of Transformations in Metals and Alloys, Part I (Pergamon Press: Oxford, 1965).
39. Y. Frenkel, The Kinetic Theory of Liquids (Oxford Univ. Press, 1946).
40. W. Lu, B. Yan, W.-h. Huang, Complex primary crystallization kinetics of amorphous Finemet alloy. J. Non-Cryst. Solids **351,** 3320-3324 (2005).
41. T. Ozawa, Kinetics of non-isothermal crystallization. Polymer **12,** 150-158 (1971).

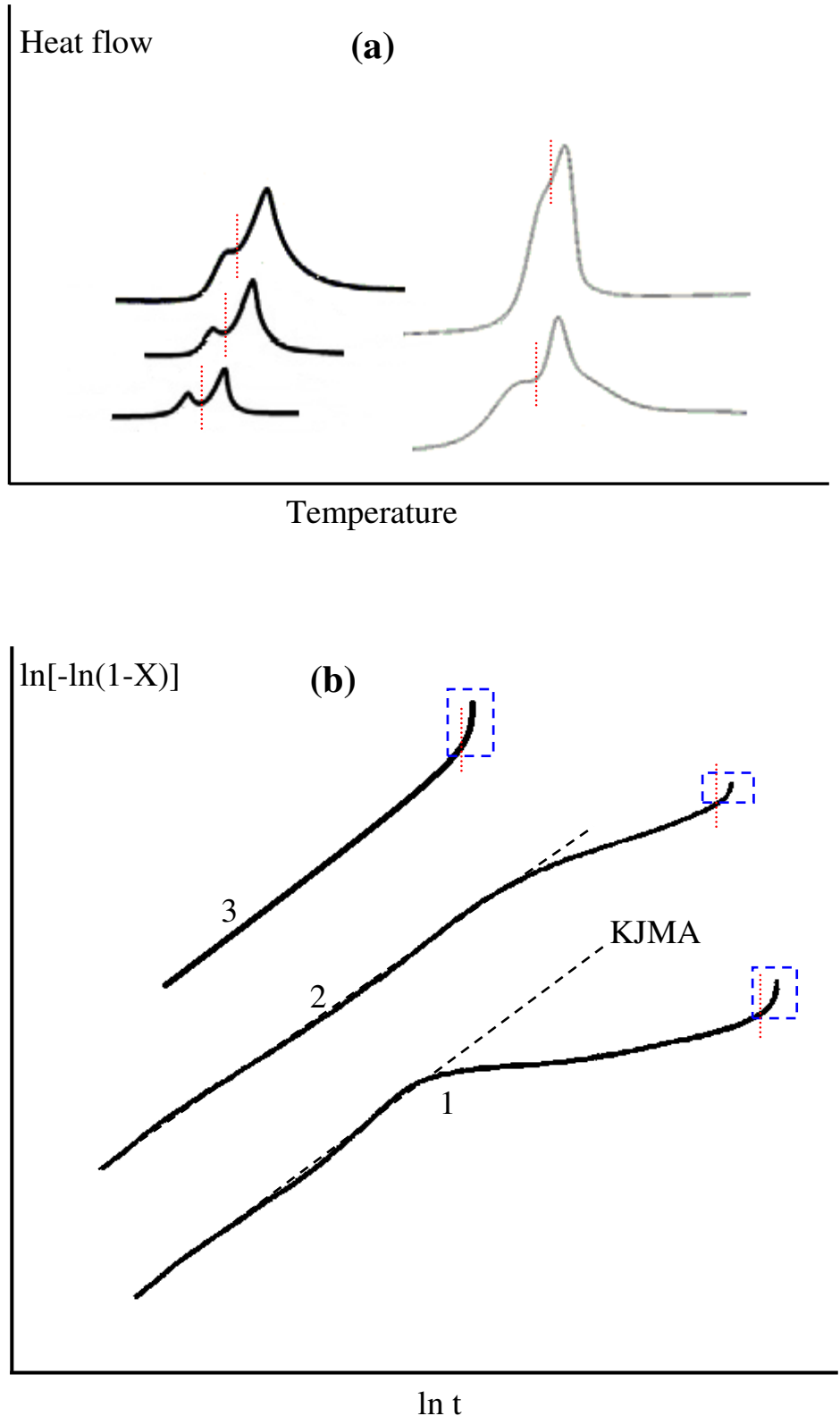

Fig.1. Examples of experimental DSC (a) [6, 7] and Avrami (b) [19] curves for the crystallization of bulk metallic glasses. BSS occurs approximately at the moments indicated by the vertical straight line. The highlighted sections of the Avrami plots in Fig. (b) correspond to the one-dimensional radial growth of the crystalline phase after BSS and thus directly indicate the GBN type of transformation.

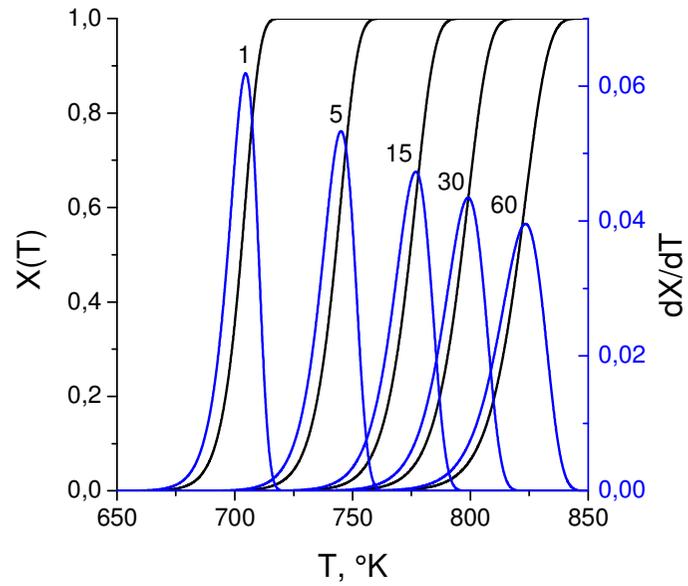

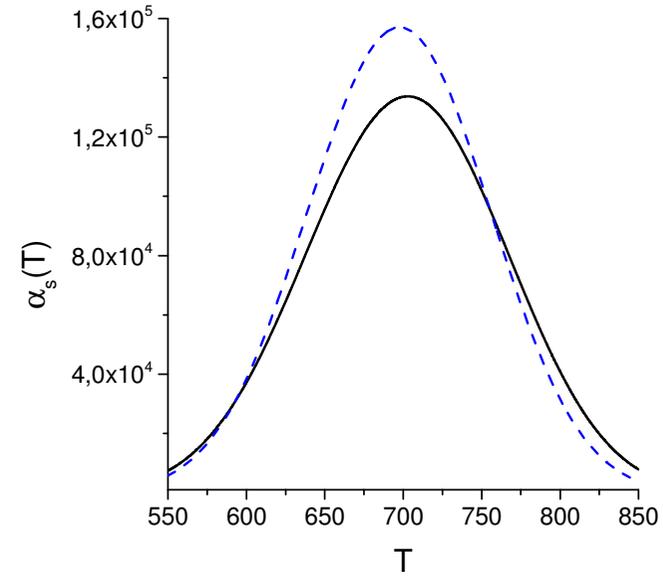

Fig.2. Volume fraction and its temperature rate of change according to the Kolmogorov equation (linear growth) for different heating rates (°K/min) indicated on the curves.

Fig. 3. Parameter $\alpha_s$ as a function of temperature for linear (solid) and diffusion (dashed) growth.

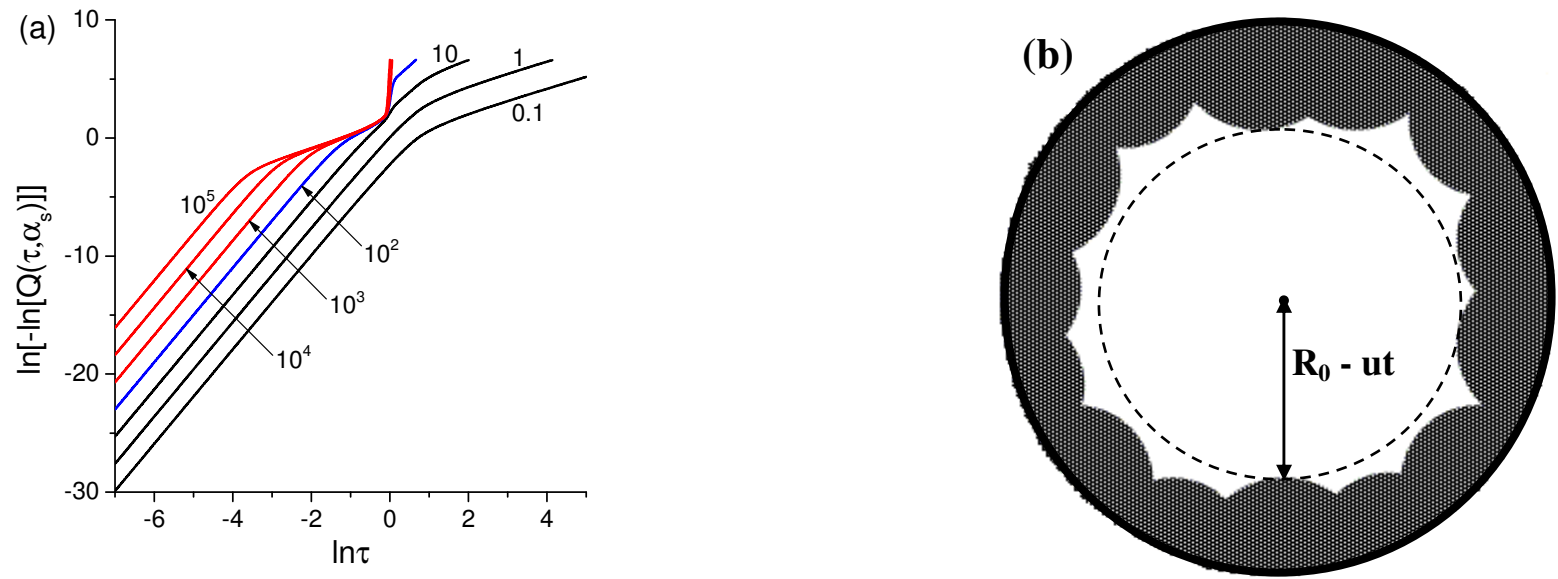

Fig. 4. (a) Avrami plots for the surface-nucleated crystallization of a spherical particle for different values of the parameter $\alpha_s$ indicated on the curves. (b) One-dimensional radial growth of the crystalline phase after saturation of the boundary of a spherical particle.

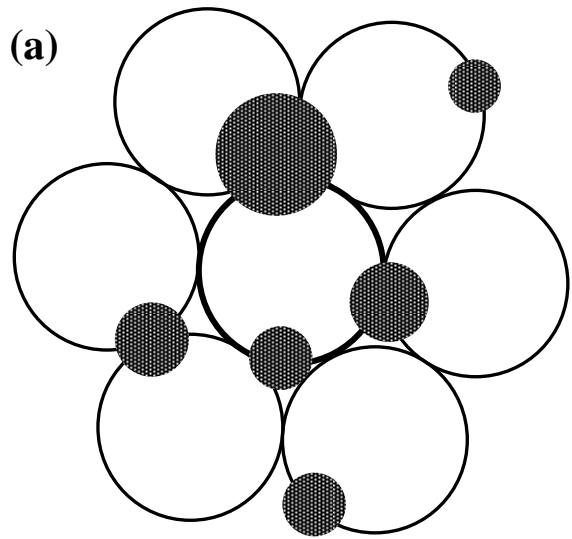 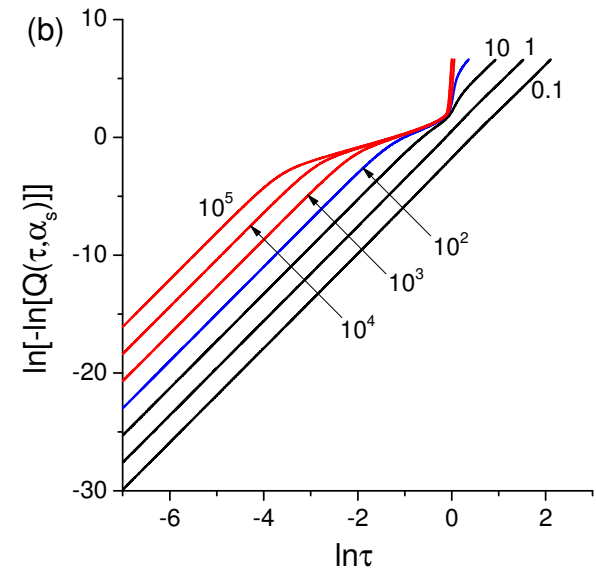

Fig. 5. (a) Grain structure of identical spherical particles. An individual particle (highlighted in the center) is transformed by nuclei formed at the boundaries of other particles as well. (b) Avrami plots for the GBN transformation of this structure for different values of the parameter $\alpha_s$ according to the VF equations of Ref. [20]. (c) $X(\tau)$ and $dX/d\tau$ curves for this structure for different values of $\alpha_s$ shown on the curves. The dotted lines correspond to one-dimensional radial growth: $1-(1-\tau)^3$ for $X(\tau)$ and $3(1-\tau)^2$ for $dX/d\tau$.

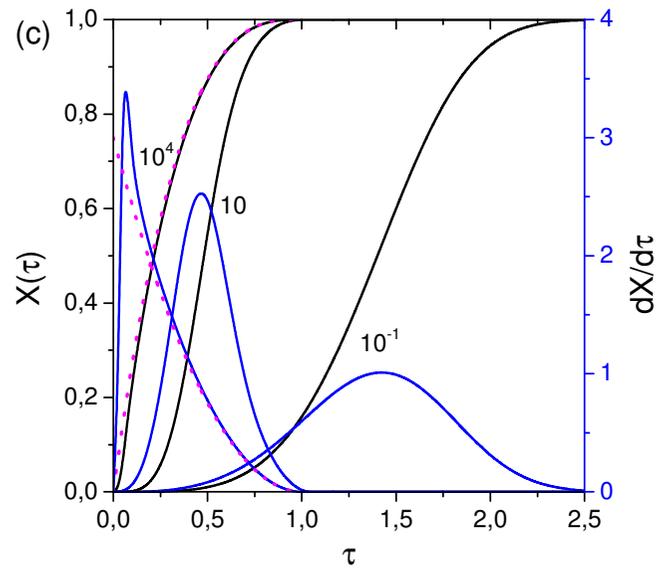

Fig. 5c.

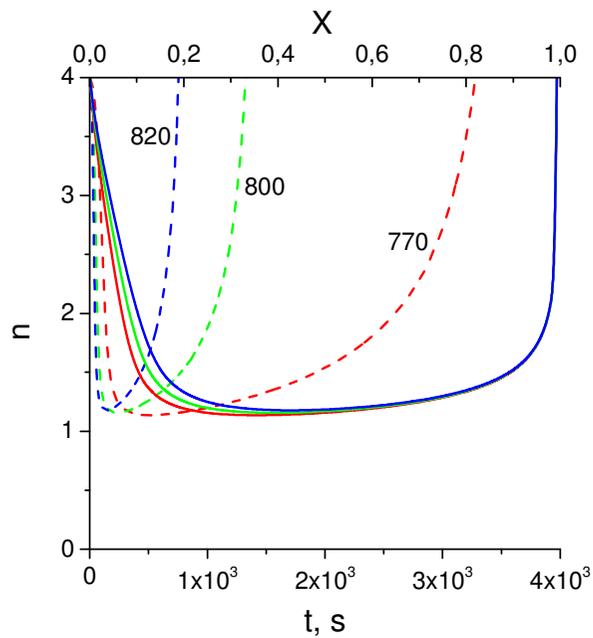

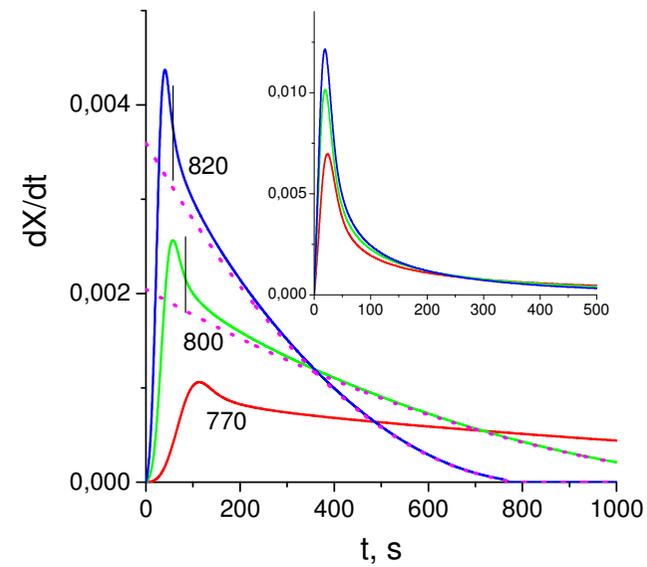

Fig. 6. Avrami exponent $n(t)$ (dashed) for the surface-nucleated crystallization of a spherical particle in the case of linear growth for three temperatures indicated on the curves and the corresponding dependences $n(X) = n(t(X))$ (solid).

Fig. 7. Isothermal DSC curves for the surface-nucleated crystallization of a spherical particle in the case of linear growth for three temperatures; the corresponding curves for the case of diffusion growth are shown in the inset. The dotted lines represent Eq. (14b). The vertical lines show the BSS moments $t_{BSS}$.

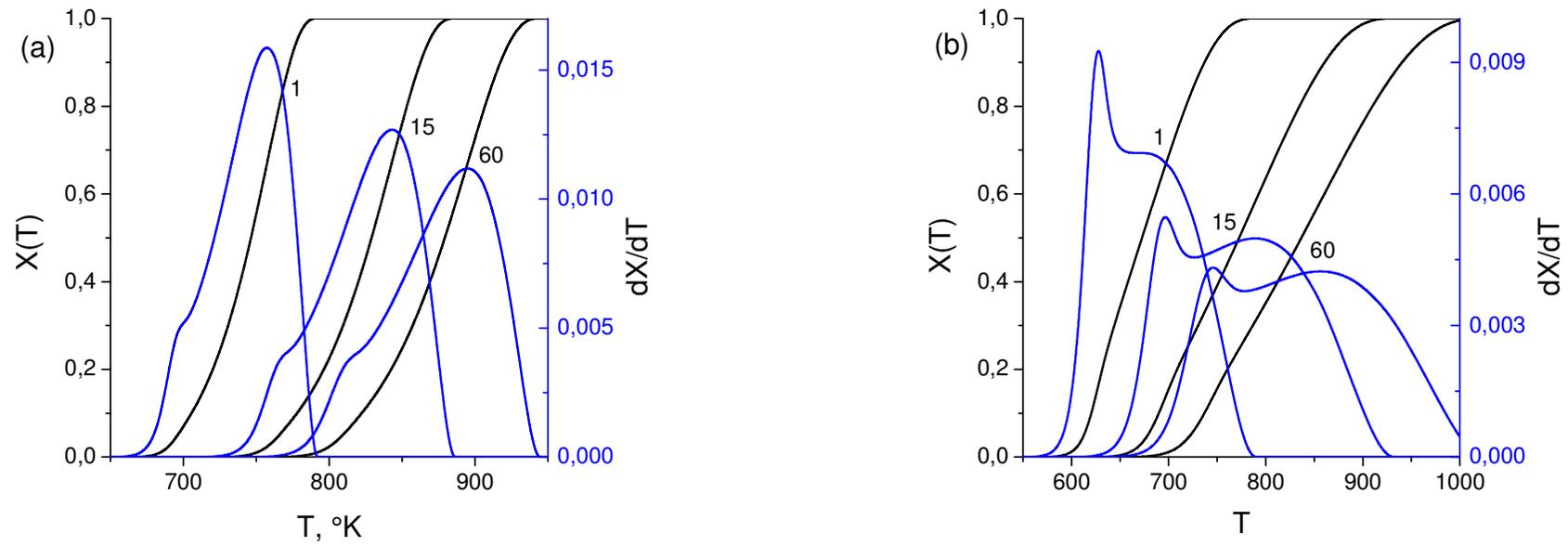

Fig. 8. Volume fraction and its temperature rate of change according to Eq. (15b) for linear (a) and diffusion (b) growth for different heating rates $\beta$ (°K/min) indicated on the curves.

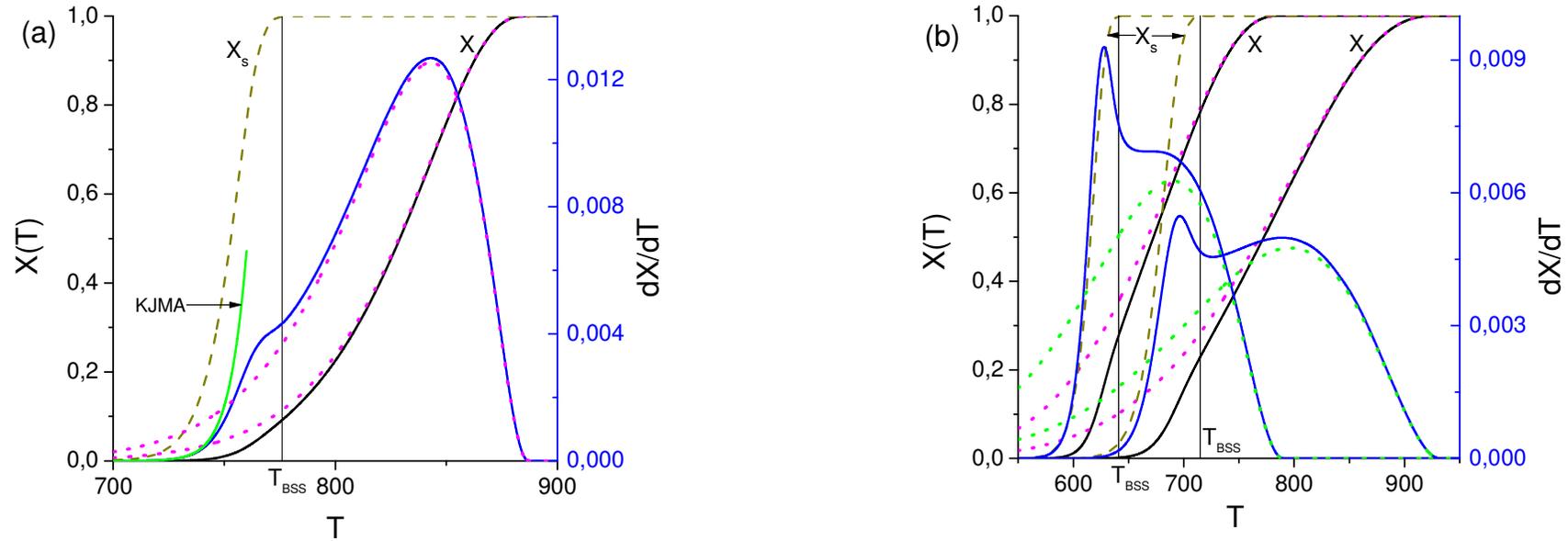

Fig. 9. (a) Part of Fig. 8a for $\beta = 15$ °K/min. The dotted lines represent Eqs. (16) and (17); the dashed line is given by Eq. (19). The vertical line shows the temperature $T_{BSS}$ at which BSS occurs. The initial (KJMA) portion of the $dX/dT$ curve is also shown. (b) Part of Fig. 8b for $\beta = 1$ and 15 °K/min. The designations are the same as in Fig. (a). The dotted lines for the $dX/dT$ curves are given by Eq. (18).